\documentstyle[twocolumn,aps,epsf]{revtex}

\begin{document}
\draft

\def \half {{1\over 2}}
\newcommand{\sss}{{\scriptscriptstyle }}
\newcommand{\sperp}{{\scriptscriptstyle\perp}}
\def\Square{{\vbox {\hrule height 0.6pt\hbox{\vrule width 0.6pt\hskip 3pt
        \vbox{\vskip 6pt}\hskip 3pt \vrule width 0.6pt}\hrule height 0.6pt}}}
\def\eV{\hbox{eV}}
\def\roughly#1{\mathrel{\raise.3ex\hbox{$#1$\kern-.75em
\lower1ex\hbox{$\sim$}}}}
\def\lsim{\roughly<}
\def\gsim{\roughly>}
\def\pref#1{(\ref{#1})}
\def\ie{{\it i.e.,}\ }
\def\eg{{\it e.g.,}\ }
\def\al{\alpha}
\def\be{\beta}
\def\ta{\tilde{a}}
\def\tb{\tilde{b}}
\def\tal{\tilde{\alpha}}
\def\tbe{\tilde{\beta}}
\input epsf.tex

\draft

\title{{\rm\small McGill-98/37, DAMTP-1998-173}\hfill {\tt\small 
hep-th/9911164}\\
\bigskip
A Naturally Small Cosmological Constant on the Brane?}

\author{C.P.~Burgess${,}^a$ R.C. Myers${}^a$ and F. Quevedo${}^b$}

\address{${}^a$ Physics Department, McGill University,
Montr\'eal, Qu\'ebec, Canada H3A 2T8.\\
${}^b$ D.A.M.T.P., Silver Street, Cambridge, CB3 9EW, England.}
\maketitle
\begin{abstract}
{
There appears to be no natural explanation for the 
cosmological constant's small size within the framework
of local relativistic field theories. We argue that
 the recently-discussed framework for which the 
observable universe is identified with a $p$-brane embedded 
within a higher-dimensional `bulk' spacetime, has special properties
that may help circumvent the obstacles to this understanding.
 This possibility
arises partly due to several unique features of the brane
proposal. These are: (1) the potential such models introduce for partially 
breaking supersymmetry, (2) the possibility
of having low-energy degrees of freedom which are not observable
to us because they are physically located on a different brane, (3) the
fundamental scale may be much smaller than the Planck scale. 
Furthermore, although the resulting cosmological constant in
the scenarios we outline need not be exactly zero, it may be
suppressed relative to the mass splittings of supermultiplets by weak
coupling constants of gravitational strength, in accord with
cosmological observations.
}
\end{abstract}
%

%
%

\section{Introduction}

There is no understanding, at present, of why the cosmological constant 
should be as small as is required to explain the enormous size of the 
observable universe. This lack of understanding is particularly vexing
from the theoretical point of view because local quantum field theories
appear to offer no way to account for the enormous disparity between
cosmological scales on one hand, and the microscopic scales of 
elementary-particle physics on the other \cite{Weinberg}.  If anything, 
this dismal situation only worsens should better data bear out present indications 
for a nonzero, but extremely tiny, cosmological constant \cite{ccnonzero},
with high-redshift supernova surveys favouring
\begin{equation}
\label{expval}
\lambda \sim (3 \times 10^{-3} \; \eV)^4 ,
\end{equation}
in units for which $\hbar = c = 1$.

The nature of the problem is this. The cosmological
constant can be considered to be the energy density 
of the vacuum, and so samples the quantum zero-point energy
density contributed by physics at any particular scale. However,
as a rule, degrees of freedom at scale $m$ contribute 
$\delta \lambda = O(m^4)$, leading to unacceptably large
results. This is true, in particular, for the theories 
which successfully describe all the well-understood physics 
associated with scales between $10^{-3}$ eV and 100 GeV. 

Supersymmetric field theories are the only known examples 
which evade this general statement, since these theories can
predict a vanishing cosmological constant even though they
involve massive particles. They can do so only if the supersymmetry
is not spontaneously broken, since in this case bosons and 
fermions precisely cancel in their contributions to the
vacuum energy. Unfortunately supersymmetry {\it must} 
be broken if it is to apply to the real world, and
the absence of superpartners for the known elementary particles
implies that the scale of this breaking must be at
least of order $m \gsim 100$ GeV. However, the resulting
failure in the bose-fermi cancellations implies a cosmological
constant which is also $O(m^4)$, and so which is much too 
large.\footnote{Although a mechanism for obtaining
supermultiplet splittings without a large cosmological constant 
has been proposed \cite{witten} for three-dimensional field
theories, its implementation in four dimensions is not clear.
See \cite{others,others2} for other related discussions of the cosmological 
constant problem.}

The purpose of this paper is to indicate a possible
way out of this dilemma, 
based on the recently much-discussed possibility that all observed 
nongravitational particles are confined to a domain-wall-like
$p$-brane which sits within a larger $(4+n)$-dimensional 
 `bulk' spacetime \cite{dimopoulos}. The choice $p=3$ gives 
the simplest picture, in which we are trapped on one of potentially 
many three-branes which sweep out a four-dimensional 
world volume within the larger-dimensional bulk space. 
Gravitational interactions,
on the other hand, are not restricted to the wall, and so are
responsible for any communication which takes place
between different branes. 
This kind of picture is actually believed to be realized within string theory,
where the branes involved can be $(p+1)$-dimensional Dirichlet 
branes  \cite{joesnotes}. We keep this particular
realization in mind, not least since it has the advantage that the resulting
brane and bulk properties are well-formulated and concrete. 

Although we do not yet have a working model,
we argue here that this framework has several new features
which may offer a way out of the usual cosmological-constant
conundrum, our purpose being to identify desirable features that
might guide explicit model building. We believe the brane 
scenario may have something to offer for understanding the 
size of the cosmological constant at both the microscopic and 
macroscopic levels. 

As viewed microscopically, it can do so because: ({\it i})
the extended supersymmetry of the bulk space permits 
the classical contributions of different branes to the cosmological 
constant to cancel, and ({\it ii}) the influence of supersymmetry
breaking can be suppressed by powers of small gravitational couplings 
because any one brane can be arranged to only {\it partially} break
these extended supersymmetries. 
Further, while similar factors arise in the mass splitting between
superpartners, our brane-world scenarios may provide an enhanced
suppression of the cosmological constant relative to these masses.

Regardless of what happens microscopically, a small cosmological
constant must also be understandable on more macroscopic scales,
right down to energies that are fractions of an eV. The brane picture 
can also help here, since it permits the existence of many low-energy
degrees of freedom which we do not see because they are trapped
on other branes to which we have no direct access. Any residual
symmetries which relate low-energy states on these other branes
to those on our own might ensure the cancellation of
contributions to $\lambda$ on scales large compared to the
inter-brane separations. 

We now elaborate on these ideas, focussing in turn on
the macroscopic and microscopic points of view.

\section{Interbrane Supersymmetry and the Macroscopic Perspective}

Any intrinsically microscopic understanding of the smallness 
of the cosmological constant cannot be complete because it leaves
open why well-understood lower-energy physics, does not 
ruin the story by contributing too strongly to $\lambda$. We
argue in this section that the brane picture
offers a new perspective for this part of the problem.

The new possibility which the brane picture introduces at low
energies is the potential it has for hiding low-energy degrees
of freedom from us. The macroscopic part of the cosmological-constant
problem states that the observed low-energy degrees of freedom
themselves induce too large a cosmological constant. However, the
difficulty vanishes if there are other low-energy degrees of
freedom about which we have no nongravitational information
which can enforce the low-energy cancellations required in $\lambda$. 

As an extreme example, suppose the branes which fill the universe 
repel one another and so in equilibrium arrange themselves into a lattice.
Further, suppose there exist residual unbroken (or very weakly
broken) discrete supersymmetry transformations which relate the 
bosons on one brane to the fermions on another.  
If such a graded lattice symmetry
were to force bosons on one brane to be degenerate with fermions
on another, and {\it vice versa}, then they can cancel in their
contributions to the effective $\lambda$ which is observed
on scales much wider than the spacings between the branes.

We do not have an explicit brane model which exhibits such a
symmetry, but as a first step we can ask whether supersymmetric
interactions can be devised in such a way as to be consistent
with the related bosons and fermions living on different branes.
Imagine, therefore, constructing a supersymmetric model consisting
of a collections of scalars, $\varphi_i$, living on one brane 
and a collection of spin-half superpartners of these bosons, $\psi_i$, 
living on another brane. We imagine coupling these fields using a
supermultiplet of `bulk' scalars and fermions, $X = \{x_a,\chi_a\}$,
which couple to both branes.
Microscopically we would have  to consider a model with supersymmetry
relating fields of different branes.
Macroscopically, at scales larger than the brane separation,
the requirement that the fields $\varphi_i$
and $\psi_i$ live on different branes amounts to asking the 
effective lagrangian to have the following additive form: 
\begin{equation}
{\cal L} = {\cal L}_b(\varphi,X) + 
{\cal L}_f(\psi,X) + {\cal L}_{\rm bulk}(X).
\end{equation}

An example of a lagrangian of this form, and which has a supersymmetry
relating the components of the supermultiplet, 
$\phi_i = \{\varphi_i, \psi_i\}$ (as well as of $X_a = 
\{x_a, \chi_a\}$), is obtained by writing a globally-supersymmetric 
Wess-Zumino model with minimal kinetic term, $K = \phi^*_i \phi_i
+ X^*_a X_a$, and superpotential 
\begin{equation}
W =  {m_{ij} \over 2} \; \phi_i \phi_j + B_i(X) \; \phi_i + C(X) ,
\end{equation}
since this implies the component interaction terms:
\begin{eqnarray}
&&{\cal L}_b^{\rm int}(\varphi,X) +
{\cal L}_{\rm bulk}^{\rm int}(X) =  - \left| m_{ij} \;
\varphi_j + B_i(x) \right|^2 \nonumber\\
&& \qquad - \left| B_{i,a}(x) \; \varphi_i
+ C_{,a}(x) \right|^2 \\
&& \qquad - \left[ \half \; \overline\chi_a \gamma_{\sss L}
\chi_b \;\Bigl( B_{i,ab}(x) \; \varphi_i + C_{,ab}(x) \Bigr)
+ c.c.\right] \nonumber\\
&&{\cal L}_f^{\rm int}(\psi,X) =  
- \half \; \overline\chi_a \gamma_{\sss L} \psi_i \; B_{i,a}(x) 
- \half \; \overline\psi_i \gamma_{\sss L} \psi_j \; m_{ij} 
+ c.c. 
\end{eqnarray}

In this model the explicit supersymmetry is not broken, so
the $\varphi_i$ and $\psi_i$ are perforce degenerate in mass.
It is this supersymmetry which keeps the vacuum energy precisely
zero. On the other hand, for sufficiently weak couplings to
the bulk fields, the observable universe on any one brane
consists of either the fields $\{\varphi_i,x_a,\chi_a\}$ or
$\{\psi_i,x_a,\chi_a\}$, which would look nonsupersymmetric
since it has in either case a mismatched number of
bosons and fermions.

In the simple toy model above, we see that the usual cosmological constant
paradox is removed in a surprising way. There are superpartners 
degenerate in mass with all of the observed particles, however,
they reside on a hidden brane, physically separated from our own,
and so cannot be directly detected.  
With this new feature, which the brane-world scenarios can provide,
the existence of these superpartners might not be in conflict with 
observations. A drawback of our toy model is that the separated
superpartners only interact through bulk fields with gravitational
strength couplings. At the moment it is not clear to us how this
mechanism could be implemented for the case where the superpartners
are charged under a gauge symmetry.
 Clearly, it would be interesting 
to find a realization of this scenario in an explicit example, 
in order to better explore its low-energy implications, as well 
as its connection with the microscopic picture, which we now describe.

\section{Extended Supersymmetry and the Microscopic Picture}

The previously-described macroscopic scenario for understanding
the smallness of the cosmological constant is pointless if the
integration over more microscopic degrees of freedom does not
also keep the cosmological constant acceptably small. We now describe
what new features the world-as-a-brane picture might add to
this part of the problem. We wish to argue that $\lambda$ may be 
very close to zero because it is protected by more than one 
supersymmetry, with not all supersymmetries directly 
broken on our own brane. 

Here we will have in mind a particular scenario involving a
system of branes having the following properties:

\noindent {\it 1. Extended Supersymmetry:} 
In the absence of the branes the bulk-space theory 
dimensionally reduces to a four-dimensional system having 
$N \ge 2$ supersymmetries.
 
\noindent {\it 2. Multiple Brane Species:}
The compactification involves several different kinds of
branes separated in the bulk. Each brane will preserve
some fraction of the $N$ bulk-space supersymmetries,
and one of these branes is imagined
to be the three-brane on which we ourselves live.

\noindent{\it 3. Partial Supersymmetry Breaking:} 
Although each of the different types of branes
preserve some of the supersymmetry, we imagine that each of the
bulk-space supersymmetries are broken by at least 
one brane. We denote by $B$ the minimum number of branes which
are required to break all of the supersymmetries. For instance,
if the bulk has two supersymmetries, while each brane respects
one supersymmetry, then two branes together could break both
supersymmetries of the bulk, and in this case $B=2$.

Thus similar to previous discussions \cite{susybranebreaking,nilles,petr}, 
supersymmetry breaking is only transmitted to our own world
from a `distant' brane. For our discussion of the cosmological
constant, however, it will be important that (at tree-level) there exist
linearly realized supersymmetries on each of the separate branes.
In ref.~\cite{petr}, Horava gives an M-theory realization of such a
scenario with two separated branes. 

\subsection{The Relevant Scales}

Microscopically, there are potentially three fundamental mass scales in the 
brane picture. The first of these is the scale, $M_b$, associated
with the inverse `width' of the brane itself. The second is the
scale, $M_s$, set by the bulk-space gravitational couplings. In the
Dirichlet-brane picture $M_b$ and $M_s$ are both set by the string scale. 
Third, there is the inverse radius, $M_c = 1/r_c$, of $n$ of the 
extra bulk-space dimensions. These dimensions are imagined to be small
compared to 
macroscopic distance scales in the four-dimensional world, but still
(much) larger than $1/M_b$ and $1/M_s$. For simplicity, we consider
$n$ extra dimensions all of roughly the same size. We disregard
any further compact dimensions with sizes of the order $1/M_s$, as they
would play no role in the following analysis. 

The scales $M_s$ and $M_c$ are not independent because 
they are related to the four-dimensional Planck mass, $M_p = 
(16 \pi G)^{-1/2} = 1.72 \times 10^{18}$ GeV (where $G$ is 
Newton's constant). For instance, in string theory this relation
is given by
\begin{equation}
\label{PlanckMc}
M_p^2 \sim e^{2\phi} \left( {M_s \over M_c} \right)^{n+2} M_c^2 ,
\end{equation}
where $e^{-\phi}$ is the closed-string coupling, and all other 
dimensionless constants are taken to be $O(1)$.

Eq.~(\ref{PlanckMc}) permits the elimination of one of $M_c$ or $M_s$ 
in favour of the other and $M_p$. A recently much-explored scenario 
daringly takes $M_s$ to be as low as the TeV scale \cite{dimopoulos}, 
in which case $M_c$ ranges from $\sim 10$ MeV (if $n=6$) to $\sim 10^{-3}$ 
eV (if $n=2$).  The requirement for a small ratio $M_c/M_s \ll 1$ 
is necessary in this picture to ensure the existence of the hierarchy 
between $M_p$ and the weak scale, $M_w$. The question of explaining 
this large hierarchy is then transferred to understanding why the 
extra dimensions are so large.

For the present purposes we must consider a different scenario, in which
$M_s$ is identified with the intermediate scale, $M_s \sim 10^{10}$ GeV.
We must do so because we envisage some supersymmetries to be breaking on distant
branes, with the news of this breaking reaching our brane only through
gravitational-strength bulk-space interactions. As we shall see, this 
generally produces supermultiplet mass splittings which are of order
$M_s^2/M_p$, which we identify with the electroweak scale, $M_w$. 
As is described elsewhere \cite{Us}, having the string scale at $10^{10}$ GeV 
has at least three other attractive features, namely: ($i$) The heirarchy
problem is solved without introducing a small value for $M_c/M_s$ as input,
since (for $n=6$) $M_w/M_p \sim M_s^2/M_p^2 
\sim e^{-2\phi} \, (M_c/M_s)^n$ is acceptably
small so long as $M_c / M_s \sim e^{-\phi} \sim 1\%$; ($ii$) The strong
CP problem can be naturally solved using the many axions found in string
theory,\footnote{Even better, there usually exist `brane axions' 
which, unlike the model independent string axion, only couple to 
the standard model gauge fields and not also to the hidden sector ones.}
since the decay constant for these axions is $M_s$, which precisely 
fits within the allowed window set by astrophysical and cosmological bounds.
($iii$) Induced neutrino masses are generically of order $m_\ell^2/M_s
\sim 10^{-1}$ eV, putting them in the right regime to account for 
neutrino-oscillations results.

Ultimately, our goal is to understand $\lambda$ as arising as a power
of the ratio $M_c/M_s$ (or, equivalently, of $M_s/M_p$). 
For these purposes $M_c$ itself is taken to be a parameter which is given, 
with no attempt made to understand the dynamics which determines 
why the small dimensions stabilize at the desired radius. The focus
instead is to understand the more difficult problem of 
how it can be that the large dimensions can remain large even once 
it is granted that the small dimensions are small.

\subsection{Cancellations at the Highest Energies}

Brane theories get off to a good start because the effective
four-dimensional vacuum energy can naturally cancel at the very
highest energies. We start with a reminder of how this cancellation
works within the D-brane context.

Imagine integrating out all of the microstructure of the branes
(such as string-scale physics for Dirichlet branes) to obtain
the (higher-dimensional) effective field theory at scales below $M_b$.
We require the low-energy effective action which governs long-distance 
gravitational effects once these higher-energy modes are integrated out. 
At scales just below $M_b$ but well above $M_c$ we have an effective
$(4+n)$-dimensional field theory of the light closed-string modes
(collectively denoted by $\Phi$, say) of the bulk space coupled 
to the light brane modes (denoted by $\Psi_b$), localized on
the various branes (which are labelled by $b$). The resulting effective 
action has the additive form:
\begin{equation}
\label{microform}
S = S_{\rm bulk}[\Phi] + \sum_b S_b[\Phi, \Psi_b].
\end{equation}

The form for the effective action at this point may be robustly
stated, because the form of the lowest-derivative terms is dictated
(in string theory) by supersymmetry. The leading terms in a low-energy 
expansion of the gravitational part of the bulk-space action are
\begin{equation}
\label{bulkaction}
S_{\rm bulk} = - \int d^{(4+n)}x \; \sqrt{-g} \; e^{2\phi}
\Bigl( M_s^{2+n} \; R + \cdots \Bigr), 
\end{equation}
where $g_{mn}$ and $R$ are the $(4+n)$-dimensional
metric and scalar curvature, respectively. $\phi$ here is
the dilaton field, which is related to the metric by supersymmetry,
and is normalized so that $e^{-\phi}$ is the closed
string coupling strength. Notice that supersymmetry precludes 
the appearance here of a bulk-space cosmological term, 
$- \int d^{(4+n)}x \; \sqrt{-g} \; \Lambda $.

The analog of eq.~(\ref{bulkaction}) for the low-energy action for 
each brane is
\begin{equation}
\label{braneaction}
S_{b} = - \int d^{(p+1)}\xi \; 
\sqrt{- \gamma} \; e^{\phi} \Bigl( \tau_b + \cdots \Bigr),
\end{equation}
where $x^m(\xi)$ is a parameterization of the `$b$th' brane's position within
the bulk space, and $\gamma_{\mu\nu} = g_{mn} \partial_\mu 
x^m \partial_\nu x^n$ is the brane's induced metric. The ellipses 
denote dependence on other fields, and on the metric's curvature, 
while the constant $\tau_b$ denotes this brane's tension.
The stability of the modes which describe the overall motion of the brane's 
centre-of-mass generally requires $\tau_b$ to be positive.

The action of eq.~(\ref{braneaction}) is also dictated by the
symmetries of the problem, being manifestly invariant with
respect to all of the supersymmetries of the underlying string theory,
provided that both brane and bulk fields are transformed. Those
supersymmetries which are unbroken by the brane
in question are realized in the usual way, with particle
states grouping into degenerate supermultiplets. The broken
supersymmetries are nonlinearly realized, however, with particles
not grouped into degenerate supermultiplets, and some brane 
fermions, $\eta$, acquiring shifts under these
transformations: $\delta \eta = \varepsilon + \cdots$ \cite{joenjim}.

A second kind of brane-like quantity which can arise in these scenarios
is a fixed surface which is not free to move. In string theory, for instance,
orientifolds are obtained by identifying points in spacetime 
which are related by a parity transformation, 
giving rise to fixed surfaces at the fixed points of these transformations. The
low-energy action acquires a contribution similar to eq.~(\ref{braneaction})
from fields evaluated on these fixed surfaces, with the
noteworthy property that the tension, $\tau_b$, for such surfaces
may be negative \cite{joesnotes,negtension}.

Imagine now integrating out all but the most slowly-varying 
bulk-field configurations, $\varphi$:
\begin{equation}
\label{resultform}
e^{i \Gamma[\varphi]} = \int {\cal D}\Phi \; e^{  iS_{\rm bulk}
[\varphi + \Phi]}  \prod_b {\cal D}\Psi_b
\; e^{ i S_b[\varphi + \Phi, \Psi_b] } .
\end{equation}
The effective cosmological constant, $\lambda$, is obtained by 
focussing on the lowest-derivative terms in the dependence
of this result on the four-dimensional metric, $g_{\mu\nu}$,
\begin{equation}
\label{4Dbulk}
\Gamma = - \int d^4x \; \sqrt{-g} \Bigl( \lambda + 
M_p^2 \; {\cal R} + \cdots \Bigr), 
\end{equation}
where ${\cal R}$ is the four-dimensional scalar curvature. 
Notice that for the purposes of identifying $\lambda$ it 
suffices to consider $g_{\mu\nu}$ infinitesimally close
to flat space.

\subsection{Microscopic Supersymmetric Cancellations}

If all of the supersymmetries of the problem are not broken
by the brane configuration of interest, then the contributions
to the cosmological constant from all of the various branes and fixed
surfaces are known to cancel quite generally. Since we will later
argue that supersymmetry can suppress the final cosmological
constant, even when broken, we first describe the supersymmetric
cancellation in more detail. 

Consider first integrating out $\Psi_b$ in the classical (tree-level) 
approximation within this low-energy theory. This corresponds to 
simply eliminating these fields using their classical
equations of motion. The cosmological constant, in this
approximation, receives a contribution from the tension
of each brane, which is typically not small for any one brane,
being generically $O(M_b^{p+1})$. Our first goal is to 
see how supersymmetry can ensure that the contributions of
the branes and fixed surfaces can robustly cancel in the cosmological
constant, without fine tuning. 

To see why this is possible, consider the case (in string theory)
of several parallel 
branes which, taken together, do not break all of the supersymmetry. 
The exact cancellation of the classical cosmological
constant is in this case related to the stability of 
these configurations due to the cancellation of the classical forces
between the branes \cite{joesnotes,DvaliTye}.

There are three steps involved in understanding
this cancellation. First notice that
the unbroken supersymmetry of any one brane configuration ensures
that the tension of any particular brane is strictly related to the value 
of the nonzero Ramond charge which it carries. It is this relation which
ensures the precise cancellation between the long-range gravitational 
attraction between parallel branes, and their long-range repulsion 
due to the force mediated by the skew-symmetric tensor gauge fields 
coupling to the Ramond charge. 

Second, the generalization of Gauss' law for the skew-tensor fields
requires that the total Ramond charge carried by all branes must sum to zero 
when the transverse dimensions are compact. This is because there 
is no place for the flux of a nonzero charge to go in a compact space. In
orientifold examples the total charge vanishes due to a cancellation 
between fixed orientifold surfaces (carrying negative Ramond charge), 
and branes (whose charges are positive). 
However, this cancellation of charges then automatically
ensures the cancellation of the negative tension of the fixed  surfaces against
the positive tension of the branes, ensuring the sum $\sum_b \tau_b = 0$. 

Third, supersymmetry ensures that this cancellation survives quantum
corrections to this classical picture. This is because the BPS nature
of the branes ensures the strict equality of their masses and Ramond charges
even at the quantum level. 

\subsection{Partial Supersymmetry Breaking}

The key question is how these arguments are modified when
the brane configuration breaks all of the supersymmetries. Although
we do not have an explicit model in hand \cite{sen}, we now wish to argue that 
the partial breaking of supersymmetries on different branes opens 
a possibility that the supersymmetric vacuum energy cancellation
might partially persist even after supersymmetry breaks. 

Recall that in the scenario we wish to consider
all of the supersymmetries of the bulk theory are broken, but that
several of the separated branes are required
to do so. In this case, because of the supersymmetry 
breaking, the forces between branes need no longer strictly cancel.
If they do not, then the branes will move until they minimize their energy. 
For instance, if the resulting forces are repulsive the branes may
try to maximize their distances from one another, perhaps
by arranging themselves into a lattice within the compact $n$
dimensions. If they instead attract one another, they may
form bound states, or simply be widely separated from one another. 
Our assumption is that the resulting stable 
configuration continues to break all of the supersymmetries.
(The possible role of slow brane motion of this type for generating
inflation was recently considered in ref.~\cite{DvaliTye}.)

Now there are two important points. First, because supersymmetry is broken,
the cosmological constant need no longer vanish. However, it cannot
become nonzero without interactions amongst the branes which
are mediated by the exchange of bulk states. Furthermore, no such contribution
to the vacuum energy is possible from a bulk exchange that is not complicated
enough to `know' that there are enough branes to break 
{\it all} of the supersymmetries.

Consider, for example, the case where our brane preserves
$N=2$ supersymmetry, and that each of these supersymmetries is broken
by a separate hidden brane. Hence diagrams such as those in figures $1a$
and $1b$ would not contribute to $\Lambda$, since they only involve
interactions between two of the branes. Their contributions
to the effective action would still preserve $N=1$ supersymmetry.
An effective cosmological constant is only induced by vacuum diagrams
involving all three branes, as illustrated in figure $1c$. 

The second point is that supermultiplet mass splittings 
are also induced by the interbrane exchange of bulk modes, however,
such masses would be induced by {\it any} such exchange (as long as the
branes involved do not preserve precisely the same supersymmetries).
Hence to leading order, interactions between only two branes will
induce mass splittings. Still, these exchanges will be suppressed
by at least two powers of the (small) bulk-space coupling, and so
we find $\delta m \sim M_b^2/M_p$, in accord with
previous supergravity analyses \cite{nilles}
(see details below). 

Hence in the example of $N=2$ supersymmetry considered above, diagrams
such as those in figures $1a$ and $1b$ (with appropriate insertions on the
observable brane) will contribute to the mass splittings. While either diagram
would individually induce a mass matrix which respects an $N=1$ supersymmetry,
these are different supersymmetries for each of the diagrams and so the total
mass matrix induced at this order would split the masses of all of the
superpartners on the observable brane. However, as discussed above,
the vacuum energy remains zero at this order, and so the suppression
of the cosmological constant may be enhanced relative to the mass splitting.

\vspace{0.25in}
\centerline{\epsfxsize=3in\epsfbox{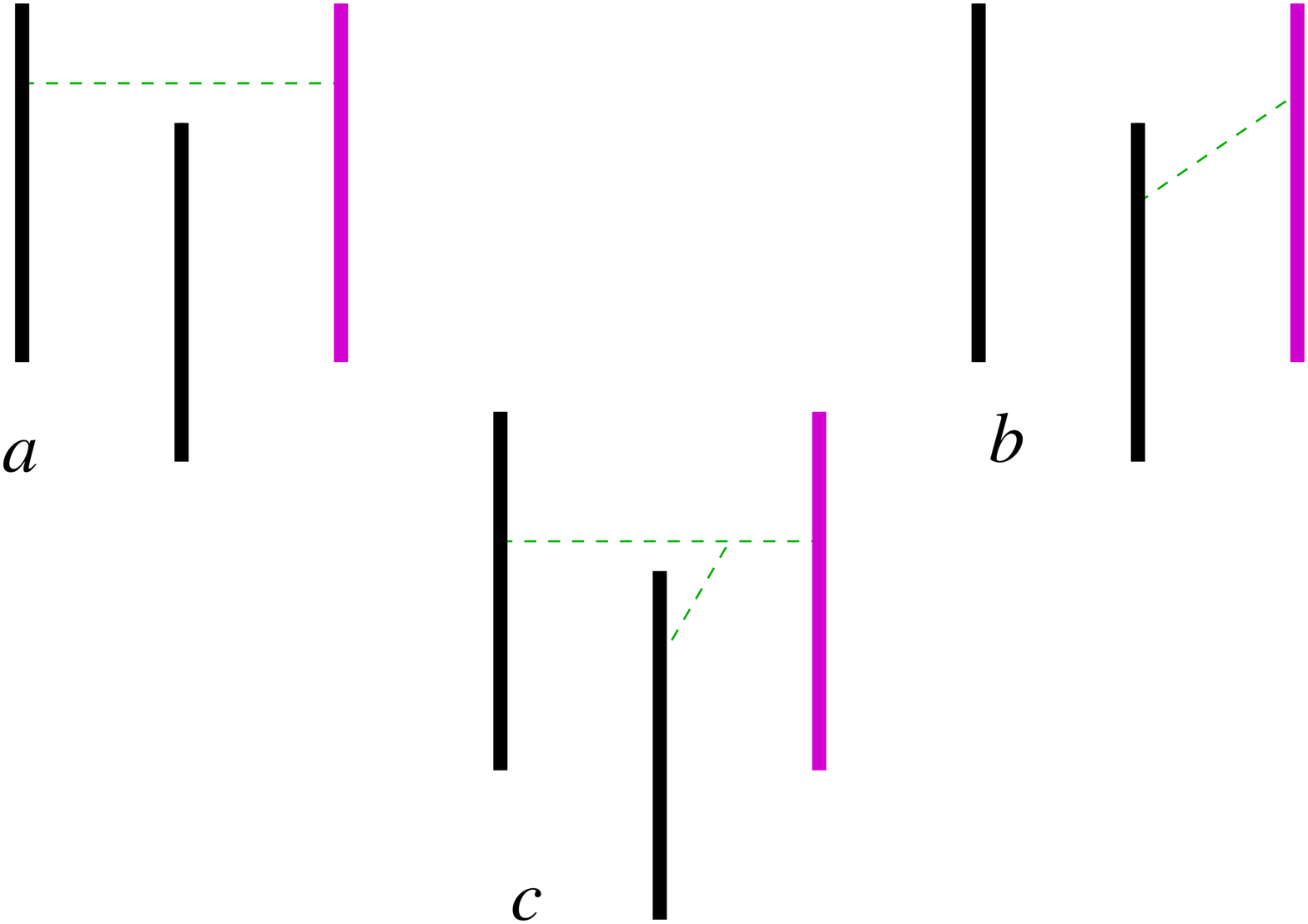}}
\medskip\noindent
{\small Figure 1:
Some tree-level graphs which potentially contribute to the cosmological
constant. Dotted lines represent the exchange of virtual bulk states
while fat lines indicate branes (the magenta 
line being the observable brane). For the $N=2$ scenario (see text),
diagram {\it c} contributes to the
cosmological constant but is suppressed by powers of the bulk coupling.
Diagrams {\it a} and {\it b} do not contribute to the cosmological constant,
but do contribute to mass splitting between superpartner fields.}
\medskip

Two comments are noteworthy here: First, this kind of suppression
because of partial supersymmetry breaking is an intrinsically braney mechanism. 
This is because, in four dimensions, it is difficult to construct models which
partially break an extended supersymmetry, as is enunciated in a
well-known apparent `no-go' theorem \cite{nogo,jon}. 
The situation is different for supersymmetric theories containing extended
objects such as domain walls or branes. In this case the no-go theorem is 
evaded \cite{joenjim,otherpartbreaking}, with a single brane typically
breaking half of the bulk-space supersymmetries and leaving the others
unbroken. More complicated configurations, involving several branes, can 
then break more of these bulk-space supersymmetries. Our second point
is that there is the potential in these scenarios to suppress the cosmological
constant relative to the supermultiplet mass splittings. This enhancement
is possible because the number of hidden branes involved in bulk exchanges
contributing to these low-energy couplings need not be the same, at
least in scenarios where more than two branes are required to break all of
the supersymmetries. We will consider this possibility in more detail
below.

\subsection{Numerology}

Although we do not have a concrete brane model in hand, 
it is nonetheless instructive to estimate the suppression which might
be expected for the cosmological constant due to these arguments, since
the results can suggest the properties to which a more detailed construction 
should aspire. Imagine, then, 
integrating out scales between $M_s$ and $M_c$, for which the universe 
effectively has $4+n$ dimensions. From a four-dimensional perspective 
this corresponds to integrating out all of the massive Kaluza-Klein 
(KK) modes whose masses are proportional to $M_c$. 

Suppose that, within
a particular brane scenario, the 
largest contribution with a nonvanishing interaction energy density requires
$a$ couplings of the bulk fields to the various branes, and $b$ self-couplings 
amongst the bulk fields themselves. On dimensional
grounds, we may estimate the resulting ($4+n$)-dimensional vacuum energy
density is of order
\begin{equation}
\label{ndcont}
\delta \Lambda \sim 
 \left[ e^{-\phi} \left({M_c \over M_s} \right)^{1+n/2} \right]^{a+b} \;
 \left( {\tau \over M_c^{p+1}} \right)^{a} \;
M_c^{4+n} ,
\end{equation}
where we put a factor of the $(4+n)$-dimensional gravitational
coupling, $\kappa=e^{-\phi}/M_s^{(1+n/2)}$
for each gravitational interaction, \ie for each of the bulk couplings
and for each coupling of bulk fields to any brane. 
For each bulk-brane coupling, we also
include a factor of the brane tension. The remaining 
powers of $M_c$ are included on dimensional grounds
since this is the scale of the physics which has been integrated out.
The corresponding four-dimensional cosmological constant induced in this
way is of order $\lambda \sim \Lambda/M_c^n$. 

At this point, we note that $a$ is at least as large as the total 
number, $B$, of branes required to break all of the supersymmetries,
but to start we also entertain the possibility that $a$ might be
larger than this.  Furthermore, in order to produce a connected
diagram, we must actually have $b\ge a-2$.

More detailed use of eq.~\pref{ndcont} requires an estimate for the
brane scale, $\tau \sim M_b^{p+1}$. In a string theory scenario involving
Dirichlet-branes, we would have $\tau \sim e^\phi M_s^{p+1}$.
In this case, eq.~\pref{ndcont} becomes
\begin{equation}
\label{4dcont}
\delta \lambda \sim 
(e^{-\phi})^{\al} \; \left( {M_c \over M_s} \right)^{\beta}  \;
M_c^{4}\ ,
\end{equation}
where the exponents are $\al=b$ and $\beta=(n/2-p)\, a+ (n/2+1)\, b$.
Let us focus on the ratio of mass scales $M_c/M_s$ (which we will assume is
smaller than the string coupling $e^{-\phi}$). We are interested in the
most dangerous graphs, which make the largest contribution
to $\delta \lambda$, and so will choose our parameters in order
to minimize the exponent $\beta$.
Here we see that we should choose the smallest possible value of $b$,
\ie $b_{min}=a-2$, which yields
\begin{equation}
\label{expone}
\beta=(n+1-p)\,a-(n+2)\ .
\end{equation}

The results beyond this point depend crucially upon the parameters
$n$ and $p$, which would be fixed for a particular scenario. There
are three mutually-exclusive alternatives:

\begin{enumerate}
\item
If $p>n+1$, the coefficient of $a$ in eq.~\pref{expone} is negative and our
estimate for $\delta\lambda$ becomes arbitrarily large as $a$ increases.
Since this means that nominally higher-order graphs contribute larger
contributions to $\delta \lambda$, we do not consider this case any further.
\item
If $p=n+1$, graphs with larger $a$ are suppressed only by powers of $e^{-\phi}$, 
and $\beta<0$, implying that higher-order graphs can {\it enhance} 
$\delta \lambda$ by an $n$-dependent power of $M_s/M_c$. For instance, with $p=3$ and
$n=2$, $\delta\lambda=(e^{-\phi})^{a-2} M_s^4$. Of course, for weak
string coupling (\ie $e^{-\phi}<1$), the largest estimate comes from
choosing the smallest possible value for $a$, which by assumption is 
the smallest number of branes which taken together break all supersymmetries,
$a_{min}=B$.
\item
The most interesting case is $p<n+1$, for which the coefficient
of $a$ in eq.~\pref{expone} is positive. Thus the largest contribution
to $\delta\lambda$ comes from setting $a$ to its minimum value of $a$,
$a_{min}=B$, the total number of branes needed to break all of the
supersymmetries. Note, however, the second term in eq.~\pref{expone}
is negative. Hence in order to produce suppression of $\delta\lambda$
with a positive $\beta$, we still need
\begin{equation}
\label{alim}
B>{n+2\over n+1-p}\ .
\end{equation}
Notice that this lower bound for $B$ is minimized when $n$ is maximized
and $p$ is minimized, so when comparing models sharing the same value
for $B$, we expect those having largest $n$ and smallest $p$ to enjoy the
largest suppression to $\delta \lambda$. 
\end{enumerate}

{}From here on we restrict ourselves to option 3, for which $p<n+1$, for
which graphs with more gravitational interactions are more strongly
suppressed by powers of $M_c/M_s$. As an example, consider $p=3$ and $n=6$,
\ie with three-branes and six extra dimensions. In this case,
we find the simple results: $\beta=4(B-2)$ and 
\begin{equation}
\label{4dconta}
\delta \lambda \sim 
\left(e^{-\phi}  {M_c^4 \over M_s^4} \right)^{B-2}  \;
M_c^{4} = \left({M_c\over M_p} \right)^{B-2} \; M_c^{4}\quad .
\end{equation}
Hence for $B>2$, we apparently find an enormous suppression of the
vacuum energy density. 

To establish the degree of the suppression more systematically,
we can compare $\delta\lambda$ with an estimate of the mass splittings
within the supermultiplets. Let us make this comparison for the case 
$p=3$, but general $n$, for which one finds
\begin{equation}
\label{4dcontb}
\delta \lambda \sim 
\left(e^{-\phi}\right)^{B-2}
\left({M_c\over M_s} \right)^{(n-2)B-(n+2)}\; M_c^{4}\quad .
\end{equation}
The condition $p=3<n+1$ forces us to focus on $n>2$, for which the coefficient
of $B$ in the exponent is positive.

In analogy to eq.~\pref{ndcont}, we may estimate the mass splittings
on dimensional grounds as
\begin{equation}
\label{ndmass}
\delta m^2 \sim 
 \left[ e^{-\phi} \left({M_c \over M_s} \right)^{1+n/2} \right]^{\ta+\tb} \;
 \left( {\tau \over M_c^{p+1}} \right)^{\ta-1} \;
M_c^{2} ,
\end{equation}
where, as above, a factor of $\kappa=e^{-\phi}/M_s^{(1+n/2)}$ appears
for each gravitational interaction. The latter includes
$\ta$ couplings of bulk fields to a brane, and $\tb$ couplings of
the bulk fields amongst themselves. For all {\it but one} of the bulk-brane
couplings, we also include a factor of the brane tension. This factor
is omitted for one of these couplings as it must involve an operator insertion,
\eg a world-volume fermion or boson bilinear, in order that this exchange
contributes to the mass splittings for supermultiplets on our brane. 
As above, the remaining powers of $M_c$ are included on dimensional grounds.

To parallel the analysis above for $\delta\lambda$, we set $\tau\sim
e^\phi M_s^{p+1}$ in eq.~\pref{ndmass} which yields
\begin{equation}
\label{4dmass}
\delta m^2 \sim 
(e^{-\phi})^{\tal} \; \left( {M_c \over M_s} \right)^{\tbe}  \;
M_c^{2}\ ,
\end{equation}
where the exponents are $\tal=\tb+1$ and $\tbe=(n/2-p)\ta+(n/2+1)\tb+p+1$.
The graphs which contribute the largest amount to $\delta m^2$ are those
having the smallest value for $\tb$, \ie $\tb_{min}=\ta-2$, which yields
\begin{equation}
\label{texpone}
\tbe=(n+1-p)\,(\ta-1)\ .
\end{equation}
Because our interest is in the case $p<n+1$, the largest
$\delta m^2$ is achieved by choosing the smallest possible $\ta$.

Now comes the key point. As we have argued in the previous
section, the smallest value for $\ta$ required to split superpartner
masses is {\it not} $B$, the number of branes needed to break {\it all} of the
supersymmetries. Rather, as argued above, supermultiplet masses can be split
by adding the contributions of several graphs, each of which breaks just
some of the supersymmetries and not all of them, although all supersymmetries
are broken once all graphs are added together. 
So we can instead set $\ta_{min}=2$,
for which $\tbe=n+1-p$. Focussing on $p=3$, we have
\begin{equation}
\label{mess5}
\delta m^2 \sim 
e^{-\phi} \; \left( {M_c \over M_s} \right)^{n-2}\;M_c^{2}
\sim e^\phi\; {M_s^4\over M_p^2}\ ,
\end{equation}
where the latter result is in accord with previous analyses 
involving the low-energy supergravity action\cite{nilles}.

Comparing the two results in eqs.~\pref{4dcontb} and \pref{mess5},
we find
\begin{equation}
\label{ratio8}
{\delta\lambda\over\delta m^4 }\sim 
(e^{-\phi})^{B-4} \; \left( {M_c \over M_s} \right)^{(n-2)B-(3n-2)}\ .
\end{equation}
Now recall that eq.~\pref{alim} gives an inequality,
$B>(n+2)/(n-2)$, which must be satisfied in order that the scale of the
vacuum energy density $\delta\lambda^{1/4}$ be suppressed relative to the
compactification scale $M_c$, down to which we are integrating out
low-energy degrees of freedom. From eq.~\pref{ratio8}, we can derive
a more interesting (and restrictive) inequality
\begin{equation}
\label{Blim}
B>{3n-2\over n-2}\ ,
\end{equation}
which must be satisfied if the
vacuum energy density, $\delta\lambda^{1/4}$, is actually suppressed relative
to the superpartner mass splitting scale $\delta m$. As an example,
consider $n=6$ (and $p=3$). Eq.~\pref{alim} yields $B>2$, while
eq.~\pref{Blim} requires $B>4$. The essential ingredient in the
separation of scales appearing in eq.~\pref{ratio8} was that the number of
branes involved in exchanges contributing to $\delta\lambda$ was $a_{min}=B$
while that number for $\delta m^2$ was $\ta_{min}=2$.

While the bulk-brane couplings contributing in
eqs.~\pref{ndcont} and \pref{ndmass} produce a suppression through the
insertion of a gravitational coupling factor $M_c/M_p$, they also tend
to enhance the result due to the brane tension factors. In particular,
for the above calculations with $\tau\sim M_s^{p+1}$, the net effect
of these couplings is to enhance the result if $p>n/2$, which will hold
in most cases. For example, if $p=3$ and $n=6$, these couplings have a
neutral effect. Hence one can think of the suppressions discussed above 
as arising from the couplings of the bulk fields to themselves, which are
necessary to `sew' the inter-brane exchanges together.

An even more interesting suppression is possible if $M_b < M_s$,
\ie $\tau<M_s^{p+1}$. For example in string theory, the branes
of interest could well be described by a  number of $D$-branes (with
positive tension)  sitting on an orientifold plane (with negative
tension) leaving a vanishing net tension. Then the leading term in 
the effective action (\ref{braneaction}) would be absent.
As a result, the bulk-brane couplings for these particular branes
also produce suppressions by the Planck scale in our estimates.
If for simplicity we assume all of the branes share this property
of a vanishing tension,
the revised estimates are found by substituting $\tau\sim M_c^{p+1}$
in eqs.~\pref{4dcont} and \pref{ndmass}, implying the absence of the 
leading order factors involving  the brane tension.
Eq.~\pref{4dcont} for the cosmological constant is then replaced by
\begin{equation}
\label{equis}
\delta \lambda \sim  \left[ e^{-\phi}
 \left({M_c \over M_s} \right)^{1+n/2} \right]^{a+b} M_c^4\;
\sim \left(M_c\over M_p\right)^{a+b}  M_c^4  .
\end{equation}
Here, as usual, the largest contribution comes from choosing the minimum values
for both $a$ and $b$: $b_{min}=a-2$ and $a_{min}=B$. With these parameters,
\begin{equation}
\label{equisb}
\delta \lambda \sim  \left[ e^{-2\phi}
 \left({M_c \over M_s} \right)^{n+2} \right]^{B-1} M_c^4\;
\sim \left(M_c\over M_p\right)^{2(B-1)}  M_c^4 
\end{equation}
and hence there is considerable suppression (relative to $M_c$)
for any number of large extra dimensions and for branes of any dimension.
Similarly, eq.~\pref{4dmass} for the mass splitting is replaced by
\begin{equation}
\label{newmass}
\delta m^2 \sim 
\left( e^{-\phi} \left({M_c \over M_s} \right)^{1+n/2} \right)^{\ta+\tb} \;
M_c^{2} 
\sim\left(M_c\over M_p\right)^{\ta+\tb}M_c^{2}\ .
\end{equation}
Again this result is maximized by choosing the smallest possible
values for the exponents, \ie $\tb_{min}=\ta-2$ and $\ta_{min}=2$.
This choice yields
\begin{equation}
\label{newmassb}
\delta m^2 \sim 
 e^{-2\phi} \left({M_c \over M_s} \right)^{n+2} \;
M_c^{2} 
\sim\left(M_c\over M_p\right)^{2}M_c^{2}\ .
\end{equation}
which is also a considerable suppression compared to $M_c$.

Within this scenario of tensionless branes, the ratio \pref{ratio8}
is then replaced with
\begin{equation}
\label{ratio8b}
{\delta \lambda\over\delta m^4} \sim  \left[ e^{-2\phi}
 \left({M_c \over M_s} \right)^{n+2} \right]^{B-3} 
\sim \left(M_c\over M_p\right)^{2(B-3)} \ .  
\end{equation}
Hence this scenario yields a remarkable suppression of the
cosmological constant relative to the scale of the supermultiplet
mass splitting for any $B>3$. Again the key ingredient in the
separation of scales produced here was that the number of
branes involved in exchanges contributing to $\delta\lambda$ was $a_{min}=B$
while that number for $\delta m^2$ was $\ta_{min}=2$.

\section{Discussion}

We see that
partial supersymmetry breaking in the world-as-a-brane framework
provides a natural mechanism by which the vacuum energy is systematically
suppressed by weak coupling constants of gravitational strength.
There is the further potential to suppress the scale of the
cosmological constant relative to that for the mass splittings
amongst supermultiplets. 

Even more tantalizing, the potential exists for naturally
generating phenomenologically interesting nonzero values for the cosmological
constant. This is because the suppression typically comes as
a power of $M_w/M_p$, and comparatively 
few powers are needed to produce an acceptably small result. 
In fact, a relatively small power like
$\lambda \sim (M_w^2/M_p)^4$ is already numerically 
roughly of order the experimental value of eq.~(\ref{expval}). 

To see what is required to reproduce such a size for $\lambda$,
we re-express the ratio \pref{ratio8b}
in terms of $\delta m \sim M_w$ and $M_p$, by using eq.~\pref{newmassb},
\begin{equation}
\label{ratio8c}
{\delta \lambda\over\delta m^4} \sim  
 \left(\delta m\over M_p\right)^{B-3} \ .  
\end{equation}
A  power like $\lambda \sim (M_w^2/M_p)^4$ is obtained in eq.~\pref{ratio8c} for 
$B=7$. In this case we have $M_s\sim 10^{13}$ GeV and $M_c\sim 10^{11}$ GeV,
(for $n/2=p=3$) therefore we have a concrete illustration of the fact
that starting from a  relatively small hierarchy 
of $M_c/M_s\sim 10^{-2}$ we can simultaneosuly obtain
$M_w/M_p\sim 10^{-16}$ (the hierarchy problem) and 
$\lambda^{1/4}/M_w \sim 10^{-16}$ (the cosmological constant problem)
\footnote{Notice the weak-scale-string
case --- $M_s \sim 1$ TeV and $n=2$, for which $M_c
\sim 10^{-3}$ eV --- is not an equally successful option (even
though $\lambda \sim M_c^4$ is the right size, as remarked
in the first article of \cite{others2}, for example) because in this
scenario {\it all} supersymmetries on our brane must be broken to
ensure particle multiplets are split by the weak scale. In this case
no supersymmetry remains to forbid $\lambda \sim (1 \hbox{TeV})^4$.}.

Similarly in the case of equation (22) both hierarchies are reproduced
if $B=10$ branes are required to break all supersymmetries, and with
branes communicating only with gravitational strength. Both again require: 
$M_s=10^{11}$ GeV, $M_c/M_s\sim 10^{-2}$.
Even though smaller values of $B$ would be more desirable,  
it is very encouraging that with such simple setting this scenario has
the potential to `explain' the large hierarchies. 
It would be of considerable interest
to explore models of these types in more detail.

Of course, as described in section II, the branes must also
provide some mechanism to maintain the suppression of $\lambda$
down to lower scales.
One of the key ingredients in the latter
separation of scales is that the number of branes involved in exchanges
contributing to $\delta\lambda$ and $\delta m^2$ need not be the same.
If a scenario like this is at work at the microscopic level,
it is ensured down to the compactification scale, $M_c$, and
then it may be perpetuated to lower energies by a macroscopic 
mechanism, such as the interbrane symmetry scenario given in
section II. If no such mechanism were to come into play,
the remaining contributions to the cosmological constant would be expected
to be unacceptably large, of order $\delta\lambda\sim M_c^4$.

Needless to say, the discussion and analysis presented here falls far short
of providing a complete solution of the cosmological constant problem.
Our primary short-coming is that we cannot offer an explicit example
which implements the mechanisms discussed above. Explicit models,
probably on similar lines as those of \cite{others,aiq}\  will be needed
in order to be more specific, and in particular, improve upon the naive
estimates provided above using simple dimensional analysis. 
Our purpose then is to point out that the brane-world scenarios
do offer a number of exciting new possibilites for suppressing
the cosmological constant, and in particular, suppressing it relative
to the superpartner mass scales. Our hope is that our observations 
may stimulate further model building along these lines.

Horava \cite{petr} has given an explicit M-theory realization of 
part of 
our scenario. However, this construction only involves two separated branes,
{\it i.e.,} boundary nine-branes. These branes are distinguished
by the appearance
of a gluino condensate on one of the branes, and not the other. Locally
each of the branes still preserves half of the supersymmetries, but
supersymmetry is still broken as a `global effect.' While this construction
involving only two branes would not give an enhanced suppression of
the cosmological constant relative to the superpartner masses,
it would still be interesting to examine the model in more detail in
the context of the questions posed here.

A number of remarks bear emphasis:

1. First and foremost, it is remarkable that the world-as-a-brane
scenario may potentially provide loopholes to the conceptual roadblocks
which have thwarted an understanding of the cosmological constant's
small size. 

2. Next, it is striking that a mechanism for producing a naturally 
small cosmological constant also points to nonzero values, possibly
in the range currently being favoured by high-$z$ supernova measurements.

3. Since individual branes tend to break half of the bulk-space
supersymmetries at once, the mechanism described here 
suggests that our own brane might have more than 
one unbroken supersymmetry, $N_{\rm us} \ge 2$. 
Of course, this is precisely the situation in which the cosmological
constant may be suppressed relative to the superpartner masses.
A careful exploration of the number of possible
unbroken supersymmetries implied by these ideas for our world 
bears further scrutiny in view of the phenomenological
problems extended supersymmetric models have, such as
the difficulty obtaining chiral fermions. We are not
daunted by these issues since we believe them to be easier
to deal with than has proven to be the case with
the cosmological-constant problem
itself.

We believe that the world-as-a-brane framework furnishes numerous
novel possibilities which may ultimately explain what keeps
the cosmological constant small without the need of fine tuning. 
Notice that the mechanism crucially involves an interplay between 
the nonlocality of brane configurations and supersymmetry.
Of course, a key test will be whether models constructed using
the mechanism we are proposing can produce a sufficiently small 
$\lambda$ while still producing a large enough splitting amongst
supermultiplets on our own brane. We believe a concrete realization 
of this idea (or variations on our theme) in terms of explicit string 
models to be well worth pursuing.

\vspace{3mm}
We acknowledge helpful discussions with Luis Ib\'a\~nez and Michael Green.
Our research is supported in part by N.S.E.R.C. of Canada and F.C.A.R. of
Qu\'ebec and  PPARC. C.B. would like to thank the Departament d'Estructura
i Constituents de la Materia of the University of Barcelona and 
D.A.M.T.P. of Cambridge University, and R.C.M. would like to thank the Institute
for Theoretical Physics at UCSB, for their kind hospitality during the final
stages of this work. Research at the ITP was supported by NSF Grant
PHY94-07194.


\begin{references}
%
\bibitem{Weinberg}
S. Weinberg, {\it Rev. Mod. Phys.} {\bf 61} (1989) 1. 
%
\bibitem{ccnonzero}
N. Bahcall, J.P. Ostriker, S. Perlmutter, P.J. Steinhardt, 
{\it Science} {\bf 284} (1999) 1481, {\tt (astro-ph/9906463)}.
%
\bibitem{witten}
E. Witten, {\it Int. J. Mod. Phys. } {\bf A10} (1995)
1247, {\tt (hep-th/9409111)}; {\it Mod. Phys. Lett.} {\bf A10} (1995) 2153,
{\tt (hep-th/9506101)}.
%
\bibitem{others}
S. Kachru and E. Silverstein, JHEP {\bf 11} (1998) 1, {\tt (hep-}
{\tt th/9808056)}; JHEP {\bf 01} (1999) 4,
{\tt (hep-th/9810129)};\\
S. Kachru, J. Kumar and E. Silverstein, {\it Phys.\ Rev.} 
{\bf D59} (1999) 106004 {\tt (hep-th/9807076)};\\
J.A. Harvey, {\it Phys. Rev.} {\bf D59} (1999) 26002, 
{\tt (hep-th/9807213)};\\
G. Shiu and H. Tye, {\it Nucl.\ Phys.} {\bf B542} (1999) 45
{\tt (hep-th/9808095)};\\
C. Angelantoj, I. Antoniadis and K. F\"orger, {\it Nucl.\ Phys.} 
{\bf B555} (1999) 116 {\tt (hep-th/9904092)}.
%
\bibitem{others2}
S. Nussinov and R. Shrock, {\it Phys. Rev.} {\bf D59} (1999) 105002
({\tt hep-ph/9811323});\\
E. Kiritsis, JHEP 10 (1999) 010 {\tt (hep-th/9906206)};\\
P.J. Steinhardt, {\it Phys. Lett.} {\bf B462} (1999) 41
{\tt (hep-th/9907080)};\\
N. Kaloper, {\tt (hep-th/9905210)}.
%
\bibitem{dimopoulos}
N. Arkani-Hamed, S. Dimopoulos and G. Dvali, 
{\it Phys. Lett.} {\bf B429} (1998) 263 ({\tt hep-ph/9803315});
Phys.\ Rev.\ {\bf D59} (1999) 086004 ({\tt hep-ph/9807344});
I.~Antoniadis, N.~Arkani-Hamed, S.~Dimopoulos and G.~Dvali,
Phys.\ Lett.\ {\bf B436} (1998) 257 ({\tt hep-ph/9804398});
P.~Horava and E.~Witten,
Nucl.\ Phys.\ {\bf B475} (1996) 94 ({\tt hep-th/9603142});
Nucl.\ Phys.\ {\bf B460} (1996) 506 ({\tt hep-th/9510209});
E.~Witten, Nucl.\ Phys.\ {\bf B471} (1996) 135
({\tt hep-th/9602070});
J. Lykken, {\it Phys. Rev.} {\bf D54} (1996) 3693
({\tt hep-th/9603133});
I.~Antoniadis, Phys.\ Lett.\ {\bf B246} (1990) 377.
%
\bibitem{joesnotes}
See, for example: J. Polchinski,  ({\tt hep-th/9611050.}),
C. Bachas, ({\tt hep-th/9806199})
and C.V. Johnson ({\tt hep-th/9812196}).
%
%
\bibitem{susybranebreaking}
T. Banks and M. Dine, {\it Nucl. Phys.} {\bf B479} (1996) 173
({\tt hep-th/9605136});\\
E. Dudas and C. Grojean, {\it Nucl. Phys.} {\bf B507} (1997) 553
({\tt hep-th/9704177});\\
I Antoniadis and M. Quir\'os , {\it Phys. Lett.} {\bf B416} 
(1998) 327 ({\tt hep-th/9707208});\\
E. Dudas, {\it Phys. Lett.} {\bf B416} (1998) 309 
({\tt hep-th/9709043});\\
T.Li, J.Lopez and D. Nanopoulos, {\it Mod. Phys. Lett.} 
{\bf A12} (1997) 2647 ({\tt hep-ph/9702237});\\
K. Choi, {\it Phys. Rev.} {\bf D56} (1997) 6588
({\tt hep-th/9706171});\\
K. Choi, H.B. Kim and C. Mu\~noz , {\it Phys. Rev.} {\bf D57} (1998)
7521 ({\tt hep-th/9711158});\\
A. Lukas, B. Ovrut and D. Waldram, 
{\it Nucl. Phys.} {\bf B532} (1998) 43 ({\tt hep-th/9710208});\\
E. Mirabelli and M. Peskin, {\it Phys. Rev.} {\bf D58} (1998) 
065002 ({\tt hep-th/9712214});\\
L. Randall and R. Sundrum, 
Nucl.\ Phys.\  {\bf B557} (1999) 79 ({\tt hep-th/9810155}).
%
\bibitem{nilles}
K.A.~Meissner, H.P.~Nilles and M.~Olechowski,
{\it Nucl.\ Phys.}  {\bf B561} (1999) 30 ({\tt hep-th/9905139});\\
H.P.~Nilles, M.~Olechowski and M.~Yamaguchi,
{\it Nucl.\ Phys.}  {\bf B530} (1998) 43 ({\tt hep-th/9801030});
{\it Phys.\ Lett.}  {\bf B415} (1997) 24 ({\tt hep-th/9707143});\\
A.~Lukas, B.A.~Ovrut and D.~Waldram,
{\it Phys.\ Rev.} {\bf D57} (1998) 7529 ({\tt hep-th/9711197});\\
Z.~Lalak and S.~Thomas,
{\it Nucl.\ Phys.}  {\bf B515} (1998) 55 ({\tt hep-th/9707223}).
%
\bibitem{petr}
P.~Horava,
Phys.\ Rev.\ {\bf D54} (1996) 7561 ({\tt hep-th/9608019}).
%
\bibitem{Us}
C.P. Burgess, L.M. Ib\'a\~nez and F. Quevedo, {\it Phys. Lett.}
{\bf B447} (1999) 257 ({\tt hep-ph/9810535});\\
See also: K. Benakli, {\it Phys. Rev.} {\bf D60} (1999)
104002 ({\tt hep-ph/9809582}).
%
%
%
\bibitem{joenjim}
J. Hughes and J. Polchinski, {\it Nucl. Phys.} {\bf B278} (1986) 147.
%
\bibitem{negtension}
S. Kachru, J. Kumar and E. Silverstein, Class.\ Quant.\ Grav.\  {\bf 17} (2000)
1139 ({\tt hep-th/9907038}). 
%
\bibitem{DvaliTye}
G. Dvali and S.H. Tye, {\it Phys. Lett.} {\bf B450} (1999)
72 ({\tt hep-ph/9812483}).
%
\bibitem{sen}
For recent reviews, with references to the literature, 
of explicit brane constructions which break supersymmetry 
see: A.~Sen, ({\tt hep-th/9904207});\\
A.~Lerda, R.~Russo, ({\tt hep-th/9905006});\\
J. Schwarz, ({\tt hep-th/9908144});\\
O. Bergman and M. Gaberdiel, Class.\ Quant.\ Grav.\  {\bf 17} (2000) 961
({\tt hep-th/9908126}).
%
\bibitem{nogo}
E. Witten, {\it Nucl. Phys.} {\bf B188} (1981) 513.
%
\bibitem{jon}
J. Bagger and J. Wess, {\it Phys. Lett.} {\bf B138} (1984) 105;\\
J. Bagger, {\it Physica} {\bf 15D} (1985) 198;\\
M. de Roo and P. Wagemans, {\it Phys.Lett.} {\bf B177} (1986) 352;\\
J.P. Gauntlett, {\it Phys.Lett.} {\bf B228} (1989) 188.
%
\bibitem{otherpartbreaking}
I. Antoniadis, H. Partouche and T.R. Taylor,
{\it Phys. Lett.} {\bf B372} (1996) 83;
S. Ferrara, L. Girardello and M. Porati, {\it Phys. Lett.}
{\bf B376} (1996) 275;
J. Bagger and A. Galperin, {\it Phys. Rev.} {\bf D55} (1997) 1091;
H. Partouche and B. Pioline, {\it Nucl. Phys. (Proc. Suppl.)} 
{\bf 56B} (1997) 322  ({\tt hep-th/9702115});\\
E. Kiritsis and Costas Kounnas, {\it Nucl. Phys.} {\bf B503} 
(1997) 117 ({\tt hep-th/9703059});\\    
S. Bellucci, E. Ivanov and S. Krivonos, {\it Phys. Lett.} {\bf B460} 
(1999) 348 ({\tt hep-th/9811244});\\
I. Antoniadis, G. D'Appollonio, E. Dudas and A. Sagnotti,
{\it Nucl. Phys.} {\bf B553} (1999) 133 ({\tt hep-th/9812118});\\
R. Altendorfer and J. Bagger, {\it Phys. Lett.} {\bf B460} (1999) 
127 ({\tt hep-th/9904213});\\ 
M. Rocek and A.A. Tseytlin, {\it Phys.Rev.} {\bf D59} (1999) 106001
({\tt hep-th/9811232}).
%
\bibitem{aiq}
I.~Antoniadis, E.~Dudas, A.~Sagnotti, {\it Phys. Lett.} {\bf B464}
(1999) 38 ({\tt hep-th/9908023});
G.~Aldazabal and A.~M.~Uranga, JHEP {\bf 10} (1999) 24 
({\tt hep-th/9908072});
G. Aldazabal, L. E. Ib\'a\~nez, F. Quevedo, 
JHEP {\bf 1} (2000) 31 ({\tt hep-th/9909172});
M. Gaberdiel, A. Sen, JHEP {\bf 11} (1999) 8 ({\tt hep-th/9908060}).
%
\bibitem{hier}
N.~Arkani-Hamed, S.~Dimopoulos and J.~March-Russell,
({\tt hep-th/9809124}).
%
\end{references}
\end{document}